\documentclass[pre,aps,preprint,floatfix,superscriptaddress,showpacs]{revtex4-1}
\usepackage{amsmath, amsthm, amssymb}
\usepackage{graphicx}

\begin{document}

\title{Pressure exerted by a grafted polymer on the limiting line of a 
semi-infinite square lattice} 

\author{Iwan Jensen}
\affiliation{ARC Centre of Excellence for Mathematics and Statistics of 
Complex Systems, Department of Mathematics and Statistics,
The University of Melbourne, VIC 3010, Australia}
\author{Wellington G. Dantas}
\affiliation{Departamento de Ci\^{e}ncias Exatas, EEIMVR,
Universidade Federal Fluminense,
27.255-125 - Volta Redonda - RJ, Brasil}
\author{Carlos M. Marques}
\affiliation{Institut Charles Sadron, Universit\'{e} de Strasbourg, 
CNRS-UPR 22, 23 rue du Loess, 67034 Strasbourg, France}
\author{J\"{u}rgen F. Stilck}
\affiliation{Instituto de F\'isica and National Institute of Science
and Technology for Complex Systems, Universidade Federal Fluminense,
Av. Litor\^{a}nea s/n, Boa Viagem, 24.210-340 - Niter\'oi - RJ, Brasil}

\date{\today}
\begin{abstract}
Using exact enumerations of self-avoiding walks (SAWs) we compute the 
inhomogeneous pressure exerted by a two-dimensional end-grafted polymer 
on the grafting line which limits a semi-infinite square lattice.  
The results for SAWs show that the asymptotic decay of the pressure as a 
function of the distance to the grafting point follows a power-law with 
an exponent similar to that of gaussian chains and is, in this sense, 
independent of excluded volume effects.
\end{abstract}


\maketitle

\section{Introduction \label{sec:intro}}

Imaging and manipulating matter at sub-micron length scales has been 
the cornerstone of nano-sciences development \cite{nano}. In Soft Matter 
systems, including those of biological relevance, the cohesive energies 
being only 
barely larger than the thermal energy $k_BT$, forces as small as a pico-Newton 
exerted over a nanometer length scale might be significant enough to induce 
structural changes. Examples can be found in the stretching  of DNA molecules 
by optical traps \cite{busta}, on the behavior of colloidal solutions under 
external fields \cite{chaikin} and on the deformations of self-assembled 
bilayers \cite{safran} to name just a few. Thus, in Soft Matter, when 
one exerts a 
localized force over a small area, precise control of the acting force 
requires not 
only a prescribed value of the total applied force but, more importantly,   
a precise pressure distribution in the contact area.

The microscopic nature of pressure has been understood since the seminal work 
of Bernoulli two and a half centuries ago: in a container, momentum is 
transferred 
by collisions from the moving particles to the walls \cite{tolman}. When the 
particle concentration is homogeneous so is the pressure. 
Strategies for localizing the pressure over a  nanometer area thus requires 
the generation of strong concentration inhomogeneities, at equivalently small 
scales. Bickel {\it et~al.} \cite{bick00,bick01} and Breidnich 
{\it et~al.} \cite{breid}
have recently realized that such 
inhomogeneities are intrinsic to entropic systems of connected particles such 
as polymer chains, and have computed the inhomogeneous pressure associated with 
end-grafted polymer chains within available analytical theories for ideal 
chains. 
Their results show that the polymer produces a local field of pressure on the 
grafting surface, with the interaction being strong at the anchoring point and 
vanishing far enough from it. Scaling arguments were also put forward 
in \cite{bick01} 
to discuss the more relevant case of real polymer chains, where excluded volume 
interactions between the different monomers need to be taken into account. 
These 
arguments suggest that the functional variation of  pressure with distance from 
the grafting point should be the same in chains with or without excluded 
volume interactions, 
albeit with different prefactors. 

\begin{figure}[h]
\begin{center}
\includegraphics[width=8.0cm]{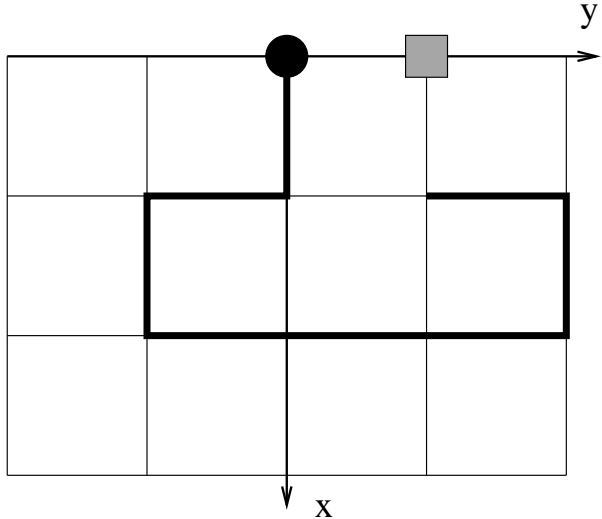}
\caption{\label{fig:model} A SAW grafted at the origin $x=y=0$ to 
a wall placed on 
the $y$ axis. If the vertex on the wall at $(0,1)$ is not excluded, the only 
possibility for the next step would be towards this vertex. If this vertex is
excluded, the SAW will end at the final point $(1,1)$.} 
\end{center}
\end{figure}

In this paper we compute the inhomogenous pressure applied to a wall by 
an end-grafted polymer with excluded volume interactions, modeled by
selfavoiding walks (SAWs) on the square lattice.
In Fig.~\ref{fig:model} 
we illustrate our model with a wall located at $x=0$.  The wall is neutral, 
in the sense that the statistical weight of a monomer placed  on the 
wall is equal to the weight of a monomer in the bulk. The length of a 
step of  the walk is equal to the lattice constant $a$, and we use 
this as the length unit.  The model is athermal, that is, all allowed 
configurations of a SAW have the same energy. 

The canonical partition function of walks with $n$ steps ($Z_n$) is equal
to the number of SAWs starting at the origin and restricted to the 
half-plane $x \ge 0$, called $c_n^{(1)}$ in \cite{b78}. The Helmholtz
free energy is given by $F_n=-k_BT \ln c_n^{(1)}$. We can estimate
the pressure exerted by the SAW at a point $(0,r)$ on the wall 
by excluding this vertex from the lattice. The excluded vertex 
is represented 
as a hatched square in Fig.~\ref{fig:model} at $r=1$.  The pressure 
$P_n(r)$ exerted at this
point is then related to the change in the free energy when the 
vertex is excluded,
$P_n a^2=-\Delta F_n$. If we call $c_n^{(1)}(r)$ the number of $n$ step SAWs
with the vertex at $(0,r)$  excluded, the dimensionless reduced pressure 
may be written as
\begin{equation}
p_n(r)=\frac{P_n(r) a^2}{k_BT}=-\ln \frac{c_n^{(1)}(r)}{c_n^{(1)}}.
\label{p}
\end{equation}
Of course we are interested in the thermodynamic limit 
$p(r)=\lim_{n \to \infty} p_n(r)$, 
so the enumeration data must be extrapolated to the infinite length limit. 
It is worth
noting that the density of monomers at the vertex $(0,r)$ is given by 
$\rho(r)=1-\lim_{n \to \infty}c_n^{(1)}(r)/c_n^{(1)}$, so that
\begin{equation} 
p(r)=-\ln [1-\rho(r)].
\label{pdens}
\end{equation}

The exact enumerations allow us to obtain precise estimates of the
pressure exerted by SAW's at small distances of the grafting point, and
we find, rather surprisingly,  that the asymptotic form of this 
pressure is well reproduced
even for these small values of $r$.
In  section \ref{sec:enum} we give some details of the computational 
enumeration 
procedure. In section \ref{sec:res} the enumeration data are analyzed 
and estimates for the pressure as a function of the distance to the 
grafting point are presented. Final discussions and conclusions may be 
found in section \ref{sec:conc}.

\section{Exact enumerations \label{sec:enum}}

The algorithm we use to enumerate SAWs on the square lattice builds on the 
pioneering work of Enting \cite{ie80} who enumerated square lattice 
self-avoiding polygons using the finite lattice method. More specifically 
our algorithm is based in large part on the one devised by Conway, Enting and 
Guttmann \cite{ceg93} for the enumeration of SAWs. The details of our
algorithm can be found in \cite{j04}.    Below we shall only briefly
outline the basics of the algorithm and describe the changes made
for the particular problem studied in this work.

The first terms in the series for the SAWs generating function can be 
calculated 
using transfer matrix techniques to count the number of SAWs in rectangles $W$ 
vertices wide and $L$ vertices long. Any SAW spanning such a rectangle 
has length 
at least $W+L-2$. By adding the contributions from all rectangles of width 
$W \leq N+1$  and length $W \leq L \leq N-W+1$ the number of SAW  
is obtained correctly up to length $N$.

\begin{figure}
\begin{center}
\includegraphics[scale=0.7]{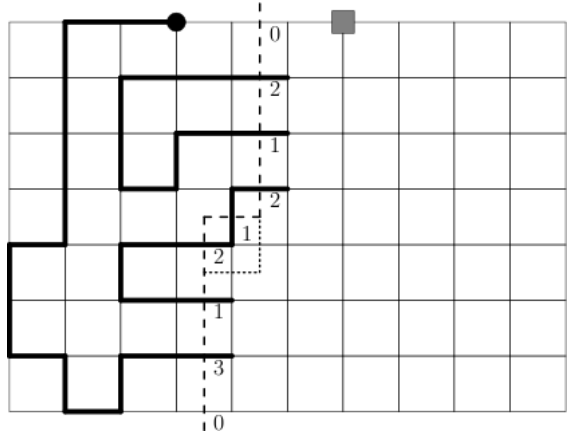}
\end{center}
\caption{\label{fig:transfer}
A snapshot of the boundary line (dashed line) during the transfer matrix (TM) 
calculation on a  strip of width 7 with $r=3$. The filled circle indicates the
grafted start-point of the SAW and the shaded box the excluded vertex.
SAWs are enumerated by successive moves of the kink in the boundary line, as 
exemplified by the position given by the dotted line, so that one 
vertex and two 
edges at a time are added to the strip. To the left of the boundary 
line we have drawn 
an example of a partially completed SAW. }
\end{figure}

The generating function for rectangles with fixed width $W$ are 
calculated using 
transfer matrix (TM) techniques. The most efficient implementation 
of the TM algorithm 
generally involves  bisecting the finite lattice with a boundary 
(this is just a line in the 
case of rectangles) and moving the boundary in such a way as to build up 
the lattice vertex by vertex as illustrated in Fig.~\ref{fig:transfer}.  
If we draw a SAW and 
then cut it by a line we observe that the partial SAW to the left of this 
line consists of a number of loops connecting two edges (we shall refer to 
these as loop ends) in the intersection, and pieces which are connected to 
only one edge (we call these free ends). The other end of the free piece is 
either the start-point or the end-point of the SAW so there are 
at most two free ends. 

Each end of a loop is assigned one of two labels 
depending on whether it is the lower end or the upper end of a loop. Each 
configuration along the boundary line can thus be represented by a set of 
edge states $\{\sigma_i\}$, where

\begin{equation}\label{eq:states}
\sigma_i  = \left\{ \begin{array}{rl}
0 &\;\;\; \mbox{empty edge},  \\ 
1 &\;\;\; \mbox{lower loop-end}, \\
2 &\;\;\; \mbox{upper loop-end}. \\
3 &\;\;\; \mbox{free end}. \\
\end{array} \right.
\end{equation}
\noindent
If we read from the bottom to the top, the configuration or signature 
$S$ along the 
intersection of the partial SAW in Fig.~\ref{fig:transfer} is 
$S=\{031212120\}$. 
Since crossings aren't permitted this encoding uniquely describes 
which loop ends are connected.

The sum over all contributing graphs is calculated as the boundary 
is moved through the lattice.  For each configuration of occupied 
or empty edges 
along the intersection we maintain a generating function $G_S$ 
for partial walks 
with signature $S$. In exact enumeration studies such as this $G_S$ is a 
truncated  polynomial  $G_S(x)$ where $x$ is conjugate to the
number of steps.  In a TM update each source  signature $S$ (before 
the boundary is moved) gives rise to a  few new target signatures  $S'$ 
(after the move of the boundary line) and $m=0, 1$ or 2 new edges are inserted 
leading to the update  $G_{S'}(x)=G_{S'}(x)+x^mG_S(x)$. Once a signature $S$ 
has been processed it can be discarded. The calculations were done using 
integer arithmetic  modulo several prime numbers with the full 
integer coefficients 
reconstructed at the end using the Chinese remainder theorem.

Some changes to the algorithm described in \cite{j04} are required in order to
enumerate the restricted SAW we study here. Grafting the SAW to the wall
can be achieved by forcing the SAW to have a free end (the start-point) 
on the top side of the rectangle. In enumerations of unrestricted SAW one 
can use symmetry to restrict the TM calculations to rectangles with  
$W\leq N/2+1$ and $L\geq W$ by counting contributions for rectangles 
with $L>W$ twice. The grafting of the start-point to the wall breaks the 
symmetry 
and we have to consider all rectangles with $W\leq N+1$. Clearly the number of 
configurations one must consider
grows with $W$. Hence one wants to minimize the length of 
the boundary line. To achieve
this the TM calculation on the set of rectangles is broken 
into two sub-sets with
$L\geq W$ and $L<W$, respectively. The calculations for the 
sub-set with $L\geq W$ 
is done as outlined above. In the calculations for the  
sub-set with $L<W$ the boundary
line is chosen to be horizontal (rather than vertical) so 
it cuts across at most  $L+1$ edges.
Alternatively, one may view the calculation for the second sub-set 
as a TM algorithm for
SAW with its start-point on the left-most border of the rectangle.

Exclusion of the vertex at distance $r$ from the starting point of 
the SAW is achieved 
by blocking this vertex so the walk can't visit the vertex.  
The actual calculation can be done 
in at least two ways.  One can simply  specify the position of 
the starting point (and $r$) 
on the upper/left  border and sum over all possible positions. This means doing
calculations for a given width $W$ many times; once for 
each position of the starting point
of the SAW.  Alternatively one can introduce `memory' into the TM algorithm. 
Specifically once we have created a configuration which inserts 
the first free end 
we `remember' that it did so.  We can flag that the free end has 
been inserted by adding 
a ghost edge to the configuration initially in state 0. 
Once the first free end is inserted the state 
of the ghost edge is changed to $1$. In the next sweep the state of 
the ghost edge
is incremented by 1. When the state of the ghost edge has 
reached the value  $r$ 
the vertex on the top border is blocked. The problem with the  
first approach is that we  
need to do many calculations for  any given rectangle. The problem with 
the second approach is that we need to  keep $r+1$ copies of most 
TM configurations thus using substantially more memory.  The 
choice will be a matter 
of whether the major computational bottle-neck 
is CPU time  or memory.  For this study we used the first approach.

In more detail the TM algorithm for the case $L\geq W$ works as follows.   
A SAW has two free ends and in the TM algorithm the first free end is forced to 
be at  the top at a distance $k$ from from the left border 
(this is the starting point of the SAW). 
We  then add a further  $r-1$ columns; in the next column the top 
vertex is forced to be empty. After this further columns are added 
up to a maximum length of  $L_{m}=N-W+1$.  This calculation 
is then repeated for 
$k=0$ to $L_{m}$ thus enumerating all possible SAWs spanning rectangles
of width exactly $W$ and length $L\geq W$. A similar calculation 
is then done with the SAW grafted to the left border and in 
each case repeated for
all $W\leq N/2$.
 
The calculation above enumerates almost all possible SAWs. 
However, we have missed 
those SAWs with two free ends in the top border where the 
end-point precedes the starting-point.
That is there is a free end in the top border at a distance 
$>r$ prior to the excluded vertex. 
We need to count such SAWs separately. The required changes 
to the algorithm are quite 
straight-forward and will not be detailed here.

We calculated the number of SAWs up to length $n=59$ for  
the unrestricted case and 
for an excluded vertex  with $r=1,\,2,\,3,\,4,\,5,\,10$, 20. 
In each case the calculation was
performed in parallel using up to 8 processors, a maximum of 
some 16GB of memory 
and using a total of under 2000 CPU hours (see \cite{j04} 
for details of the parallel algorithm). 
We needed 3 primes to represent each series correctly and 
the calculations for all the primes 
were done in a single run.

\section{Analysis and results \label{sec:res}}

\begin{figure}
\begin{center}
\includegraphics[width=10.0cm]{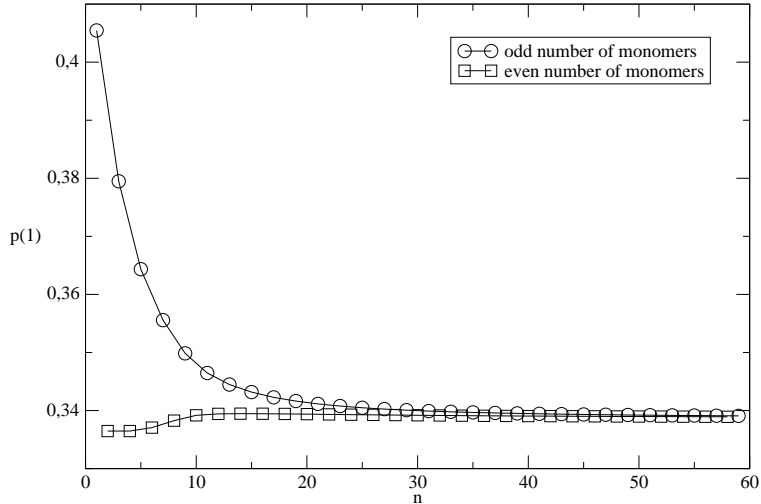}
\caption{\label{fig:peuler} Pressure $p_n(r)$ for $r=1$, calculated with the 
enumeration data for $c_n^{(1)}$ and $c_n^{(1)}(1)$ using expression (\ref{p}).} 
\end{center}
\end{figure}

In tables \ref{tab:cn1} and \ref{tab:cn11}, we have listed the 
results for the enumerations of
self-avoiding walks without additional restrictions, $c_n^{(1)}$, and 
walks which are not
allowed to occupy the vertex $(0,1)$ of the wall, $c_n^{(1)}(1)$. 
If we calculate
the pressures directly, we notice a parity effect, as seen
in the results presented in Fig.~\ref{fig:peuler}. This effect is 
related to an unphysical singularity in the generating function of the counts
$c_n^{(1)}$, $G(x)=\sum_{n=0}^\infty c_n^{(1)}x^n$. Besides the physical singularity
at $x=x_c=1/\mu$, where $\mu$ is the connective constant, there is another 
singularity at $x=-1/\mu$ \cite{b78}. This point will be discussed 
in more detail 
below, and more precise estimates for the pressures at several distances from 
the grafting point will be provided.

\begin{table}
\caption{Number  of walks in the half-plane $c_n^{(1)}$. \label{tab:cn1}}
\begin{tabular}{c@{\hspace{4mm}}l@{\hspace{8mm}}c@{\hspace{4mm}}l@{\hspace{8mm}}c@{\hspace{4mm}}l}
\hline
\hline
$n$ &  $c_n^{(1)}$ & $n$ &  $c_n^{(1)}$  & $n$ &  $c_n^{(1)}$ \\
\hline
1 & 3 & 21 & 681552747 & 41 & 176707555110156095 \\ 
2 & 7 & 22 & 1793492411 & 42 & 465629874801142259 \\ 
3 & 19 & 23 & 4725856129 & 43 & 1227318029107006037 \\ 
4 & 49 & 24 & 12439233695 & 44 & 3234212894649555857 \\ 
5 & 131 & 25 & 32778031159 & 45 & 8525055738741918835 \\ 
6 & 339 & 26 & 86295460555 & 46 & 22466322857670716727 \\ 
7 & 899 & 27 & 227399388019 & 47 & 59220537922987286933 \\ 
8 & 2345 & 28 & 598784536563 & 48 & 156073168859898607113 \\ 
9 & 6199 & 29 & 1577923781445 & 49 & 411414632591966686887 \\ 
10 & 16225 & 30 & 4155578176581 & 50 & 1084313600069268939547 \\ 
11 & 42811 & 31 & 10951205039221 & 51 & 2858360190045390998925 \\ 
12 & 112285 & 32 & 28844438356929 & 52 & 7533725151809823220637 \\ 
13 & 296051 & 33 & 76016486583763 & 53 & 19860118923927104821817 \\ 
14 & 777411 & 34 & 200242023748929 & 54 & 52346889766180530489735 \\ 
15 & 2049025 & 35 & 527735162655901 & 55 & 137997896899080793506959 \\ 
16 & 5384855 & 36 & 1390287671021273 & 56 & 363744527134008049572583 \\ 
17 & 14190509 & 37 & 3664208598233159 & 57 & 958930393586321187515995 \\ 
18 & 37313977 & 38 & 9653950752700371 & 58 & 2527696511232818406275131 \\ 
19 & 98324565 & 39 & 25444550692827111 & 59 &  6663833305674862002802763   \\ 
20 & 258654441 & 40 &    67042749110884297     \\ 
\hline
\hline
\end{tabular}
\end{table}

\begin{table}
\caption{Number of restricted walks in the half-plane $c_n^{(1)}(1)$. 
\label{tab:cn11}}
\begin{tabular}{c@{\hspace{4mm}}l@{\hspace{8mm}}c@{\hspace{4mm}}l@{\hspace{8mm}}c@{\hspace{4mm}}l}
\hline
\hline
$n$ &  $c_n^{(1)}(1)$ & $n$ &  $c_n^{(1)}(1)$ & $n$ &  $c_n^{(1)}(1)$\\
\hline
1 & 2 & 21 & 484553893 & 41 & 125845983216200025 \\ 
2 & 5 & 22 & 1277403184 & 42 & 331741159147128245 \\ 
3 & 13 & 23 & 3361118347 & 43 & 874112388226242422 \\ 
4 & 35 & 24 & 8860136085 & 44 & 2304278197456842952 \\ 
5 & 91 & 25 & 23319106552 & 45 & 6071977423574762560 \\ 
6 & 242 & 26 & 61468398004 & 46 & 16006835327039914244 \\ 
7 & 630 & 27 & 161814936995 & 47 & 42181825940070651834 \\ 
8 & 1672 & 28 & 426530787110 & 48 & 111200914189945767681 \\ 
9 & 4369 & 29 & 1123043680259 & 49 & 293056004233059019257 \\ 
10 & 11558 & 30 & 2960232320818 & 50 & 772575890795109134325 \\ 
11 & 30275 & 31 & 7795418415398 & 51 & 2036121996024316003415 \\ 
12 & 79967 & 32 & 20548006324647 & 52 & 5367866589569286706072 \\ 
13 & 209779 & 33 & 54117914172220 & 53 & 14147607361624429924807 \\ 
14 & 553634 & 34 & 142651034798697 & 54 & 37298221266819312654286 \\ 
15 & 1453801 & 35 & 375747632401071 & 55 & 98307470253293931954939 \\ 
16 & 3834878 & 36 & 990456507011029 & 56 & 259178303320281122974230 \\ 
17 & 10077384 & 37 & 2609158017850105 & 57 & 683144867659867533730505 \\ 
18 & 26574366 & 38 & 6877742334133961 & 58 & 1801074652042354959971779 \\ 
19 & 69870615 & 39 & 18119629209950641 & 59 & 4747450605648675761162683 \\ 
20 & 184216886 & 40 & 47764129557587369 \\ 
 \hline
\hline
\end{tabular}
\end{table}

\subsection{Critical points and exponents \label{sec:crpexp}}

The critical behaviour of a polymer grafted to a surface is well 
established \cite{dbl93}.
It has been proved that the connective constant of grafted walks
equals that of unrestricted walks \cite{w75}.  The associated 
generating function
has a dominant singularity at $x=x_c=1/\mu$ 
 \begin{equation}\label{eq:gf}
G(x) = \sum_{n} c_n^{(1)}x^n \sim A(1-\mu x)^{-\gamma_1},
\end{equation}
where $\gamma_1=61/64$ is a known \cite{ds86,d89} critical exponent.
Besides the physical singularity  there is another 
singularity at $x=x_-=-x_c$ \cite{gw78,b78}. 

We have analysed the series using differential approximants \cite{g89}.  We calculate
many individual approximants and obtain estimates for the critical points
and exponents by  an averaging procedure described in chapter 8 of
reference \cite{g09}. Here and elsewhere uncertainties on estimates  from
differential apprimants  was obtained from the spread among the
various  approximants as detailed in  \cite{g09}. 
The results for  unrestricted grafted SAWs are listed in Table~\ref{tab:crpexp}
under $r=0$. We also list estimates for the cases $r=1,\, 2,\, 5$ and 10..
From these estimates it is clear that all the series have the same critical
behaviour. That is a dominant singularity at $x=x_c$ with exponent
$-\gamma_1=-61/64$ and a non-physical singularity at $x=x_-=-x_c$ 
with a critical exponent consistent with the exact value $\gamma_-=3/2$.

The critical behavior can be established more rigorously from a simple
combinatorial argument. The number of walks $c_n^{(1)}(r)$ with the point at
$(0,r)$ excluded is clearly less than the number of unrestricted walks  
$c_n^{(1)}$. 
On the other hand if we attach a single vertical step to the grafting 
point of an 
unrestricted walk we get a walk  which does not touch the surface at all and
hence these walks are a subset of $c_n^{(1)}(r)$. This establishes the inequality
\begin{equation}\label{eq:cnr}
c_{n-1}^{(1)} \leq c_n^{(1)}(r) \leq c_n^{(1)},
\end{equation}
and hence shows that  up to amplitudes the asymptotic behaviors 
of these sequences are identical.

\begin{table}[htdp]
\caption{\label{tab:crpexp} Estimates of the critical points and exponents for 
SAWs with an excluded vertex a distance $r$ from the origin ($r=0$ is 
the unrestricted case).
The estimates were obtained from third order approximants with $L$ being the
degree of the inhomogenous polynomial.}
\begin{center}
\begin{tabular}{llllll}  \hline \hline
$r$ & $L$ & \multicolumn{1}{c}{$x_c$} &  \multicolumn{1}{c}{$\gamma$} & 
 \multicolumn{1}{c}{$x_-$} & \multicolumn{1}{c}{$\gamma_-$} \\ \hline
0    &  0  & 0.379052260(64)  & 0.953097(70)  & -0.3790526(38)  & 1.5002(19)\\
0    &  4  &  0.379052241(20)& 0.953072(17)& -0.3790492(30)& 1.5023(13) \\
0    &  8  & 0.379052243(14)& 0.953071(15) & -0.3790498(21)& 1.5016(12) \\ \hline
1  &  0  & 0.3790522582(30)& 0.9530884(24)    & -0.3790425(97)& 1.5074(74) \\
1  &  4  &  0.3790522575(38)& 0.9530879(30)    & -0.379030(26)& 1.523(29) \\
1  &  8  &    0.379052257(11)& 0.953090(14)&  -0.379058(16)& 1.4988(69) \\  \hline
2  & 0 &   0.379052292(16)& 0.953123(13)     & -0.3790511(33)& 1.5011(24) \\
2  & 4 &  0.379052276(12)& 0.9531115(97)&  -0.3790478(89)& 1.5036(60) \\
2  & 8 &   0.379052306(26)& 0.953135(20)&  -0.379057(21)& 1.498(20) \\  \hline
5  & 0 & 0.37905218(21)& 0.95304(17)    & -0.379114(61)& 1.457(37) \\ 
5  & 4 & 0.37905225(31)& 0.95313(24)&   -0.379099(40)& 1.467(29) \\
5  & 8 & 0.37905226(29)& 0.95313(25)&  -0.379074(31)& 1.482(20) \\  \hline
10 & 0 & 0.3790483(12)& 0.9494(12)   & -0.379230(55)& 1.369(32)  \\
10 & 4 &  0.3790493(40)& 0.9503(32)&  -0.379237(29)& 1.369(14) \\
10 & 8 &  0.3790508(22)& 0.9514(14) &  -0.379246(91)& 1.365(54) \\
\hline \hline
\end{tabular}
\end{center}

\end{table}%

\subsection{Pressure \label{sec:pres}}

Having established the critical behaviour of the series we can now 
turn to the determination
of the pressure exerted by the polymer on the surface.  Since all 
the series have the same dominant
critical behaviour it follows from (\ref{p}) that the pressure is 
given by the ratio of the critical amplitudes. 

One way of estimating the amplitudes is by a direct fit to an 
assumed asymptotic form. Here we assume that the asymptotic behaviour of 
our series is similar to that  of unrestricted SAW \cite{Caracciolo}. The asymptotic 
analysis of \cite{Caracciolo}  was very thorough and clearly established that the 
leading non-analytic correction-to-scaling  exponent has the value 3/2 
(there are also analytic, {\it i.e}., integer valued corrections to scaling). 
We repeated some of the steps in this analysis with the same result for the 
leading non-analytic correction-to-scaling exponent. Naturally there may be further
non-analytic correction-to-scaling  exponents with values $>3/2$, but these
would be impossible to detect numerically with any degree of certainty.
 So here we assume that the physical singularity has a leading 
correction-to-scaling exponent of 1 followed by further half-integer corrections while we 
assume only integer corrections
at the non-physical singularity. We thus fit the coefficients 
to the asymptotic form
\begin{equation}\label{eq:asymp}
c_n^{(1)}(r)= \mu^n\left[ n^{\gamma_1-1} \left( A(r) +\sum_{j=2}a_j(r)/n^{j/2}\right)+ 
(-1)^nn^{-\gamma_- -1}\sum_{k=0}b_k(r)/n^k\right].
\end{equation}
In the fits we use the extremely accurate estimate  $\mu=2.63815853035(2)$ 
obtained  from an analysis of the series for
self-avoiding polygons on the square lattice \cite{cj11} and the conjectured 
exact values $\gamma_1=61/64$ and $\gamma_-=3/2$.
That is we take a sub-sequence of terms
$\{c_n^{(1)}(r),c_{n-1}^{(1)}(r),\ldots,c_{n-2m-1}^{(1)}(r)\}$, plug 
into the formula above 
taking $m$ terms from both the $a_j$ and $b_k$ sums, and solve the 
$2m$ linear equations to obtain estimates for the amplitudes.

\begin{figure}
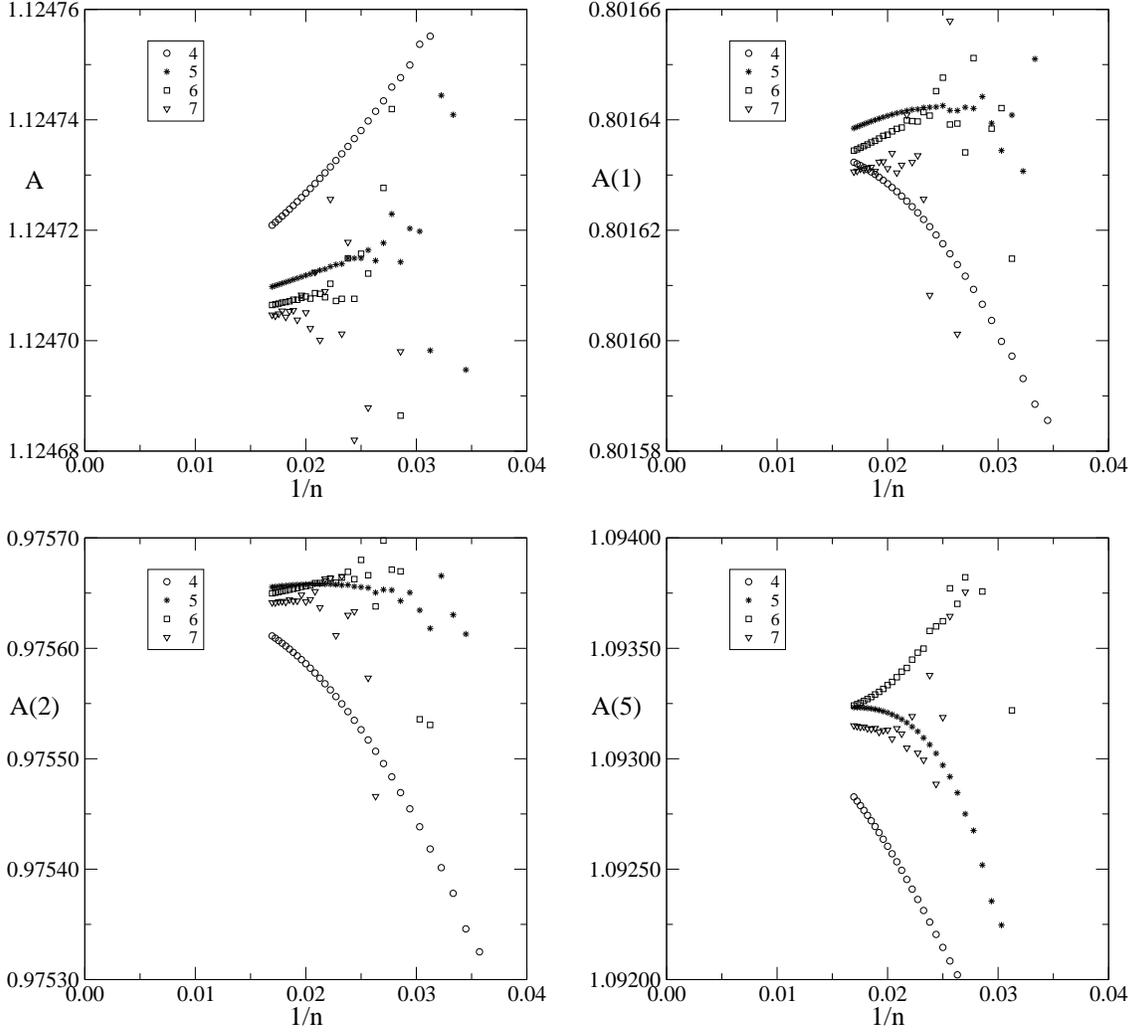

\begin{picture}(440,400)
\put(0,200){\includegraphics[scale=0.4]{Ampl_nr}}
\put(220,200){\includegraphics[scale=0.4]{Ampl_r1}}
\put(0,0){\includegraphics[scale=0.4]{Ampl_r2}}
\put(220,0){\includegraphics[scale=0.4]{Ampl_r5}}
\end{picture}
\caption{\label{fig:ampl} Estimates for the leading amplitudes 
obtained by fitting to the asymptotic
form \protect{(\ref{eq:asymp})} plotted against $1/n$ while truncating
the asymptotic expansion 
after 4 to 7 terms. 
}
\end{figure}

It is then advantageous to plot estimates for the leading amplitude 
$A(r)$  against $1/n$ for several values of $m$ as done in Fig.~\ref{fig:ampl}. 
The behaviour of the estimates for the leading amplitudes shown in  this figure 
supports that (\ref{eq:asymp}) is a very good approximation to the true asymptotic form. 
In particular note that the slope becomes very flat as $n$ is increased and 
decreases as the number of terms $m$ included in the  fit is increased.
From these plots we
estimate $A=1.124705(5)$, $A(1)=0.801625(5)$, $A(2)=0.97564(2)$ and 
$A(5)=1.09325(10)$, where the uncertainty is  a conservative value
chosen to include most of the extrapolations from Fig.~\ref{fig:ampl}.

The amplitude ratios $A(r)/A$,  and hence the pressure, can also be
estimated by direct extrapolation 
of the relevant quotient sequence, using a method due to Owczarek et
al. \cite{o94}: 
Given a sequence $\{a_n\}$ defined for $n \ge 1$, assumed to converge
to a limit $a_{\infty}$ 
with corrections of the form $a_n \sim a_{\infty}(1 + b/n + \ldots)$,
we first construct a new sequence $\{p_n\}$ 
defined by $p_n = \prod_{m=1}^n a_m$. We then analyse the
corresponding  generating function 
$$P(x)=\sum p_n x^n \sim (1 - a_{\infty} x)^{-(1+b)}.$$
Estimates for $a_{\infty}$ and the parameter $b$ can then be obtained
from differential approximants, 
that is $a_{\infty}$ is just the reciprocal of the first singularity
on the positive real axis of $P(x)$. 
In our case we study the sequence of ratios
$a_n(r)=c_n^{(1)}(r)/c_n^{(1)}$, which has the required 
asymptotic form. Using the same type of differential approximant 
method outlined above
we find that $A/A(1)=1.4030218(5)$, which is entirely consistent with the 
estimate
$A/A(1)=1.403030(15)$ obtained using the amplitude estimates from the
direct fitting procedure.  

\begin{figure}
 \includegraphics[scale=0.6]{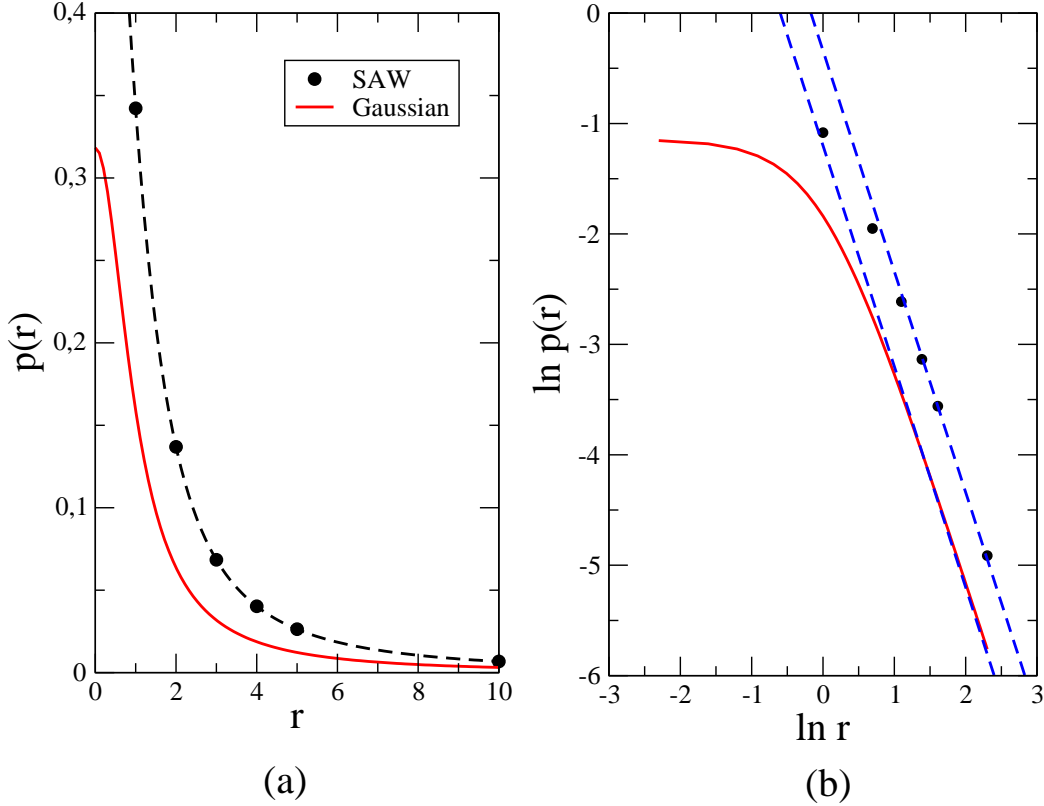} 
\caption{\label{fig:press}  (a) The pressure $p(r)$ exerted by a polymer
  on a surface at a distance $r$ 
from the grafting point. Data are for polymers modelled as SAWs or
Gaussian chains. The dashed line are a $1/r^2$ fit.
(b) Both data have the same $1/r^2$ scaling form,
even for values close to $r=2$. The dashed lines are guide lines with 
slope equal to $-2$.}
\end{figure}

Next we compare these results for the pressure with the ones for gaussian chains
as expressed in equation~(4) in \cite{bick00}. That expression is for polymers in a  
three-dimensional half-space confined by a two-dimensional wall, and corresponds to  
finite values of the radius of gyration. If the expression is generalized to the 
$d$-dimensional  case and restricted to the limit of infinite chains, where the radius
of gyration diverges, the result is:  
\begin{equation}\label{eq:Gauss}
p_G(r)=\frac{P_G(r)a^d}{k_BT}=\frac{\Gamma(d/2)}{\pi^{d/2}}
\frac{1}{(r^2+1)^{d/2}},  
\end{equation}
where we recall that $r$ is dimensionless, measured in units of the
lattice constant $a$. 
In  table \ref{tab:press} we have listed the estimated pressures for
SAWs and the pressures
obtained for Gaussian chains in $d=2$, on the semi-infinite square lattice. 
In Fig.~\ref{fig:press} we
have plotted the pressure for polymers modelled as SAWs  and as
Gaussian chains. 
In this figure the dashed line represents a decay in pressure with the
same asymptotic 
form, $\propto 1/(r^2+1)$, as the Gaussian chain but normalised so the
curve passes through 
the SAWs data point for $r=10$. Quite clearly the SAWs data is well
represented by this form 
even for small distances $r>2$. For $r=20$ the SAWs data was
indistinguishable from zero pressure.

\begin{table}
\begin{tabular}{ccc}
\hline
\hline
$r$ & $p(r)-SAWs$ & $p(r)$-gaussian \\
\hline
1 & 0.33863 & 0.15915 \\
2 & 0.14218 & 0.06366 \\
3 & 0.07334 & 0.03183 \\
4 & 0.04347 & 0.01872 \\
5 & 0.02844 & 0.01224 \\
10 & 0.00735 & 0.00315 \\
\hline
\hline
\end{tabular}
\caption{\label{tab:press} Pressure at a distance $r$ from the
  grafting point for SAWs and Gaussian chains.}
\end{table}
 
\section{Final discussions and conclusion \label{sec:conc}}
\begin{figure}
 \includegraphics[scale=0.5]{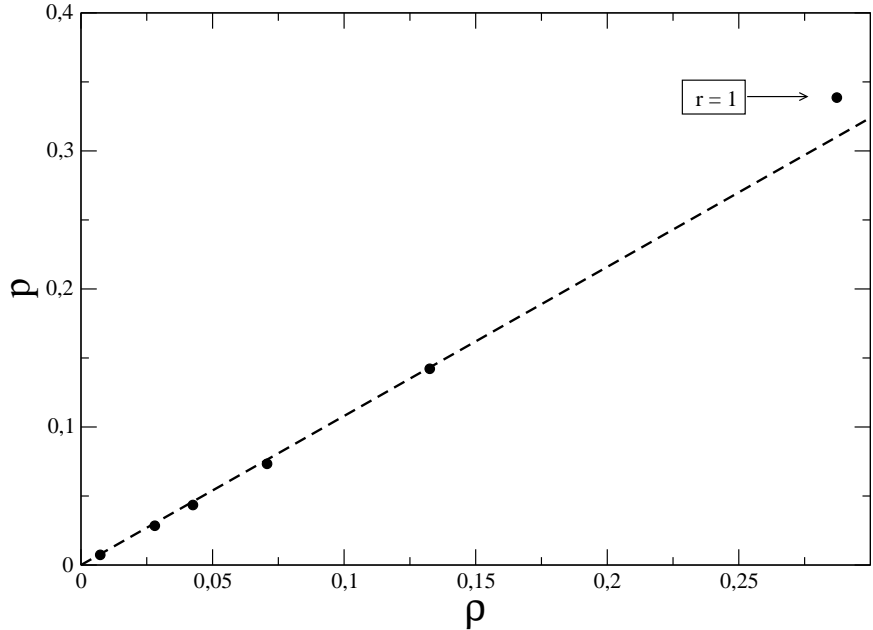} 
\caption{\label{fig:prerho}  
Relation between pressure and concentration of monomers near
to the wall at a distance $r$ from the grafting point. 
For $r>1$, a linear relationship is observed.}
\end{figure}

Since our model is athermal and discrete, it is not really possible
to compare our results with those obtained for the gaussian chain. 
However, as was already mentioned by Bickel {\it{et al.}}~\cite{bick01}, 
the excluded volume interactions should not change the scaling form 
of the pressure. Fig.~\ref{fig:press}(b) clearly shows a $1/r^2$ decay 
of the pressure, even for small distances.
According to Bickel {\it{et al.}} \cite{bick01}, this similarity is due
to the fact that the pressure and the monomer concentration
in the vicinity of the wall are linearly related. On the other hand,
it seems that the concentration is not affected by the molecular
details or by the differences between chain models. In our case,
despite the fact that $\rho(r)$ and $p(r)$ are related by a logarithmic
relation, as shown in expression (\ref{pdens}), we have for $r\gg 1$
a small concentration leading to a linear relation between those
quantities. Actually, even for $r\sim 2$, we can observe a linear
dependence, as shown in  Fig.~\ref{fig:prerho}.

Since the grafted chain is in mechanical equilibrium, the force
${\cal F}$ applied to the walk at the grafting point, which is in the 
negative $x$ direction in Fig.~\ref{fig:model},  should be
equal to the sum of the forces applied by the wall at other contact
points, which are in the positive $x$ direction. Thus, the dimensionless 
force is given by:
\begin{equation}
f=\frac{{\cal F}a}{k_BT}=2\sum_{r=1}^\infty p(r).
\label{ftotal}
\end{equation}
For gaussian chains, integrating equation~(\ref{eq:Gauss}), we find
$f_G=1$. For SAWs, we may estimate the force summing the results for
$r=1,2,\ldots,5$ and obtaining the remaining contributions
($r=6,7,\ldots,\infty$) using the
asymptotic result $p(r) \approx A_p/(r^2+1)$ where $A_p \approx
0.74235$ was estimated using the result of $p(r)$ for $r=10$. The
result of this calculation is $f_{SAW} \approx 1.533$, larger than the one
for gaussian chains. As mentioned above, it does not seem straightforward 
to compare the
two models, since a gaussian chain is a mass-spring model and therefore
it is, unlike SAWs, 
not athermal. We may also mention that if $p(r)$ for SAWs
is extended to real values of $r$ using a numerical interpolation procedure
and the data for gaussian chains are rescaled so the areas below both curves
are the same, the difference between the curves is quite small, the maximum
being close to the origin and of order $10^{-3}$. Due to the limited precision 
of the estimates for SAWs and to the expected small dependency of the 
results on the interpolation procedure we will not present these results
here, but we found that in general the 
rescaled results for the pressure of gaussian chains are larger than 
the pressures for SAWs at small values of $r$, but the inverse situation 
is found for larger distances. This net effect  may be understood if 
we recall that the pressure is a monotonically growing function of 
the local density at the wall (Eq.~(\ref{pdens})) and that 
the effect of the excluded 
volume interactions should be a slower decay of this density with
the distance from the grafting point, as compared to approximations
where this interaction is neglected.

It is of some interest to obtain the total force applied to the
chain at the grafting point for ideal chains, modeled by random
walks on the semi-infinite square lattice. This force may be 
calculated considering the shift of the grafting point by one
lattice unit in the positive $x$ direction in Fig.~\ref{fig:model}. 
The change in 
free energy under this operation will be proportional to the force.
This calculations should lead to the same result 
of the ones above, where the force was obtained summing over the
pressures at all other sites of the wall besides the origin, since
the total force applied on the chain has to vanish. 

Let us start by briefly reviewing the calculation of the number of
random walks on a half-plane of the square lattice. If we call
$c_n(\vec{\rho})$ the number of $n$-steps random walks on a
square lattice starting at the origin and ending at the
point $\vec{\rho}=x{\bf i}+y{\bf j}$, the number of RWs on the
half-plane $x \ge 0$ may be 
calculated by placing an absorbing wall at $x=-1$, so that any walk
reaching the wall is annihilated. This may be accomplished by using an
image walker, starting at the reflection point of the origin with
respect to the 
wall and ending at $\vec{\rho}$. We will place the starting point
of the random walk at $(s,0)$, where $s=0$ corresponds to walks starting
at the origin. In this case the
image walker starting point will be at
$\vec{\rho}_0=-(s+2){\bf i}$, with 
distances measured in units of the lattice 
constant $a$. The number of walks confined to the $x \ge 0$ half plane
is given by \cite{rg04}
\begin{equation}
c^{(1)}_n(\vec{\rho},s)=c_n(\vec{\rho})-
c_n(\vec{\rho}+(2+s){\bf i}).
\label{eq:iw}
\end{equation}
Since we are interested in the large $n$ limit, we may use the
gaussian approximation for the number of walks
\begin{equation}
c_n(\vec{\rho})=\frac{4^n}{n\pi}
\exp\left(-\frac{|\vec{\rho}|^2}{n}\right).
\end{equation}
For the half-plane we get
\begin{equation}
c_n^{(1)}(\vec{\rho},s)=\frac{4^n}{n\pi}
\left[\exp\left(-\frac{|\vec{\rho}|^2}{n}\right)-
\exp\left(-\frac{|\vec{\rho}+(2+s){\bf i}|^2}{n}\right)\right]
\end{equation}
To obtain the total number of walks, we integrate this expression over
the final point $\vec{\rho}$
\begin{equation}
c_n^{(1)}(s)=\int_0^{\infty}dx\int_{-\infty}^{\infty}dy\;
c_n^{(1)}(\vec{\rho},s).
\end{equation}
The result is
\begin{equation}
c_n^{(1)}(s)=\frac{4^n}{\sqrt{\pi}}
\int_{-s/\sqrt{n}}^{(2+s)/\sqrt{n}}e^{-x^2}dx,
\end{equation}
for $n \gg s$, we have the asymptotic behavior
\begin{equation}
c_n^{(1)}(s)=4^n\frac{2(s+1)}{\sqrt{n\pi}},
\end{equation}
which has the expected scaling form (\ref{eq:gf}), with exponent
$\gamma=1/2$ and amplitude $A=2(s+1)/\sqrt{\pi}$. The change in
free energy between the cases with $s=0$ and $s=1$ is therefore
given by $-k_BT\ln 2$, so that the force applied to the polymer 
by the wall at the grafting point will be $f_{RW}=\ln 2 
\approx 0.6931$, which is lower than the forces obtained for 
gaussian chains  and estimated for SAWs.

It should be mentioned that for SAWs the sum
of the pressures corresponding to two distances $p(r_i)+p(r_j)$ is
always smaller (for finite $|r_i-r_j|$) than $-\Delta
F(r_i,r_j)/(k_BT)$, where $\Delta F(r_i,r_j)$ is the change in free
energy when both cells, at $r_i$ and $r_j$ are excluded. In other
words, an effective attractive interaction exists between the two
excluded cells, so that the free energy decreases as the cells
approach each other. This effect is due to walks in the unrestricted case
which visit both excluded cells, and are therefore not counted in
either $c_n^{(1)}(r_1)$ or  $c_n^{(1)}(r_2)$.
The total force $f_{SAW}^\prime$, resulting from the simultaneous exclusion 
of all cells besides the
one at the grafting point r = 0, must thus be smaller than the 
force $f_{SAW}$ defined in equation (\ref{ftotal}).
It is easy to find, since the
number of SAWs with $n$ steps $d_n^{(1)}$ in this case is given by
$d_n^{(1)}=1+c_{n-1}^{(1)}$, that for a given value of $n$ the force at the
grafting point will be $f^\prime_{n,SAW}=-\ln(d_n^{(1)}/c_n^{(1)})$. For large $n$,
we get $f^\prime_{SAW}=\ln \mu \approx 0.9701$, smaller than 
$f_{SAW}=1.533$, as expected.

Finally, we should also stress that although the pressure applied by the 
SAWs and by the gaussian chains display a similar power-law behavior,
other possible walks on the lattice might lead to different results. 
Recently the pressure exerted by directed walks starting at the origin 
on the limiting line of a semi-infinite square lattice was obtained \cite{rp12}.
In the limit of large directed walks the asymptotic decay of the 
pressure with the distance to the grafting point also follows a power 
law, albeit with an exponent smaller than the one obtained here for 
SAWs and gaussian chains.

\begin{acknowledgments} 
We would like to thank Neal Madras for useful comments.
The computations for this work was supported by an award under the
Merit Allocation Scheme on the NCI National Facility  at the 
Australian National University. We also made use of the 
computational facilities of the Victorian Partnership for Advanced Computing.
IJ was supported under the Australian Research Council's Discovery Projects 
funding scheme by the grants DP0770705 and DP1201593. JFS acknowledges 
financial support by the brazilian agency CNPq.
\end{acknowledgments}

\end{document}